\documentclass[a4paper]{jpconf}
\usepackage{graphicx}
\usepackage{mathrsfs,amsmath,amssymb,amsfonts}
\begin{document}
\title{Pygmies, giants, and skins as laboratory constraints on the
equation of state of neutron-rich matter}

\author{J. Piekarewicz}

\address{Department of Physics, Florida State University,
                  Tallahassee, FL 32306-4350, USA}

\ead{jpiekarewicz@fsu.edu}

\begin{abstract}
 Laboratory experiments sensitive to the density dependence of the
 symmetry energy may place stringent constraints on the equation
 of state of neutron-rich matter and, thus, on the structure,
 dynamics, and composition of neutron stars. Understanding the 
 equation of state of neutron-rich matter is a central goal of nuclear 
 physics that cuts across a variety of disciplines. In this contribution 
 I focus on how laboratory experiments on neutron skins and on 
 both Pygmy and Giant resonances can help us elucidate the structure 
 of neutron stars.
 \end{abstract}

\section{Introduction}
\label{introduction}
\begin{figure}[h]
\begin{center}
 \includegraphics[height=5cm]{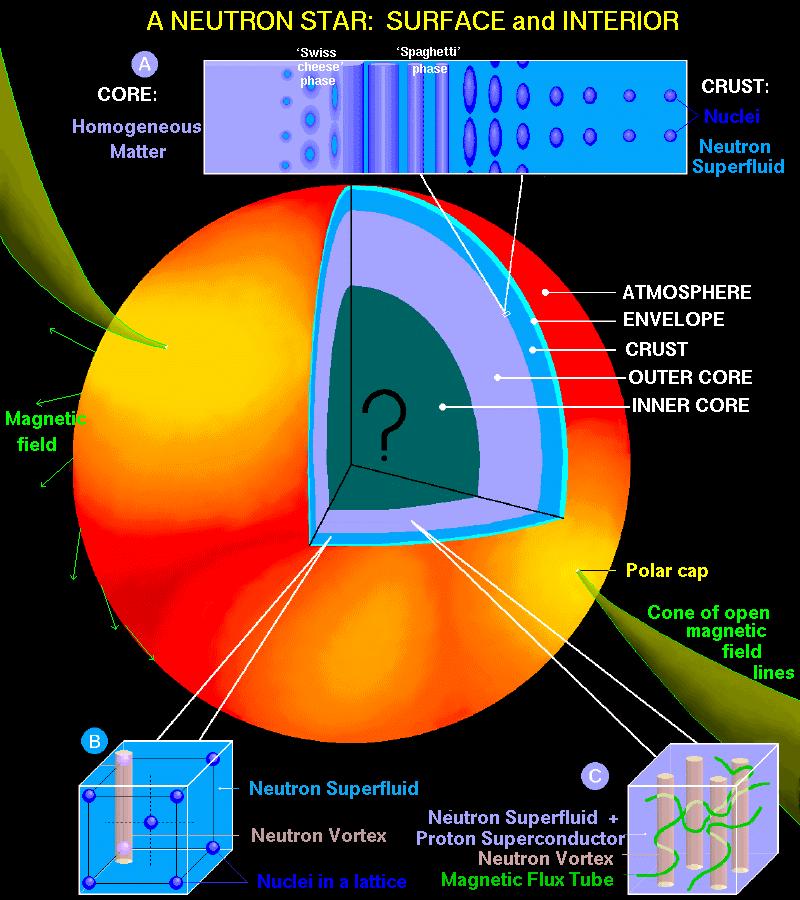}
 \includegraphics[height=5cm]{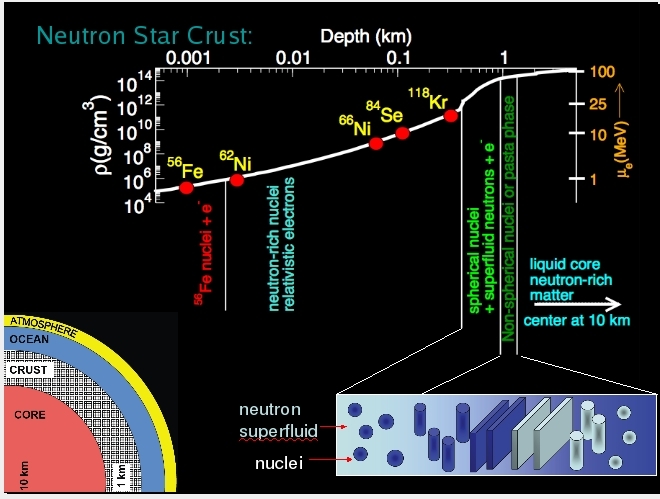}
 \vspace{-0.2cm}
 \caption{A scientifically-accurate rendition of the structure 
               and phases of a neutron star (courtesy of Dany Page) 
               and the composition of the stellar crust (courtesy of 
               Sanjay Reddy).}
 \label{Fig1}
\end{center}
\end{figure}

One of the central questions framing the recent report by The Committee 
on the Assessment of and Outlook for Nuclear Physics is {\sl ``How does 
subatomic matter organize itself?''}\,\cite{national2012Nuclear}. Remarkably, 
most of the fascinating phases that are predicted to emerge in this subatomic 
domain can not be probed under normal laboratory conditions. However, such 
novel states of matter become stable in the interior of neutron stars by virtue
of the enormous gravitational fields. The fascinating phases predicted to exist 
in the crust of neutron stars, such as Coulomb crystals of neutron-rich nuclei 
and nuclear pasta, are within the purview of the Facility for Rare Isotope Beams 
(FRIB) which has as one of its science goals to provide an understanding of 
matter in the crust of neutron stars\,\cite{2007LongRangePlan}. FRIB is of 
relevance to neutron-star structure because at sub-saturation densities the 
uniform neutron-rich matter residing in the stellar core becomes unstable 
against cluster formation. That is, at these sub-saturation densities the 
separation between nucleons increases to such an extent that it becomes
energetically favorable for the system to segregate into regions of normal 
density ({\sl i.e.,} nuclear clusters) embedded in a dilute, likely superfluid, 
neutron vapor. This {\sl ``clustering instability"} signals the transition from the 
uniform liquid core to the non-uniform stellar crust (see Fig.\,\ref{Fig1}).

The outer stellar crust of relevance to the FRIB program is comprised of a 
Coulomb lattice of neutron-rich nuclei immersed in a uniform electron 
gas\,\cite{Baym:1971pw,Ruester:2005fm,RocaMaza:2008ja,RocaMaza:2011pk}). 
At the lowest densities of the outer crust, the Coulomb lattice is formed from 
${}^{56}$Fe nuclei. However, as the density increases---and given that the electronic 
Fermi energy increases rapidly with density---it becomes energetically favorable for 
electrons to capture into protons leading to the formation of a Coulomb crystal of 
progressively more neutron-rich nuclei; a progression that starts with ${}^{56}$Fe 
and is predicted to terminate with the exotic, neutron-rich nucleus ${}^{118}$Kr 
(see Fig.\,\ref{Fig1}). 

Eventually, the neutron-proton asymmetry becomes too large for the nuclei to 
bind any more neutrons and the excess neutrons go into the formation of a 
dilute neutron vapor. However, at inner-crust densities, distance scales that 
were well separated in both the crystalline phase (where the long-range Coulomb 
interaction dominates) and in the uniform phase (where the short-range strong 
interaction dominates) become comparable. This gives rise to {\sl ``Coulomb frustration''}, 
a phenomenon characterized by the formation of a myriad of complex structures 
radically different in topology yet extremely close in energy.  Given that these
complex structures---collectively referred to as {\sl ``nuclear pasta''}---are very close 
in energy, it has been speculated that the transition from the highly ordered crystal 
to the uniform phase must proceed through a series of changes in the dimensionality 
and topology of these structures\,\cite{Ravenhall:1983uh,Hashimoto:1984}. Moreover,
due to the preponderance of low-energy states, frustrated systems display an interesting 
and unique low-energy dynamics that has been studied using a variety of techniques 
including numerical simulations\,\cite{Horowitz:2004yf,Horowitz:2004pv,
Horowitz:2005zb,Watanabe:2003xu,Watanabe:2004tr, Watanabe:2009vi}.

FRIB will not be the only experimental program of direct relevance to the
structure of neutron stars. Indeed, the {\sl Lead Radius EXperiment} (``PREX'') 
at the Jefferson Laboratory measures the neutron radius of ${}^{208}$Pb. When
combining this purely electroweak result with the accurately known charge radius 
of ${}^{208}$Pb, one obtains its {\sl ``neutron skin''}---the difference between the 
root-mean-square neutron and proton radii. As we shall see later, the neutron skin 
correlates strongly to the pressure of pure neutron matter at saturation density 
which, in turn, correlates strongly to the neutron-star radius.  

\section{Formalism}
\label{formalism} 

\begin{figure}[h]
\begin{center}
 \includegraphics[height=5.75cm]{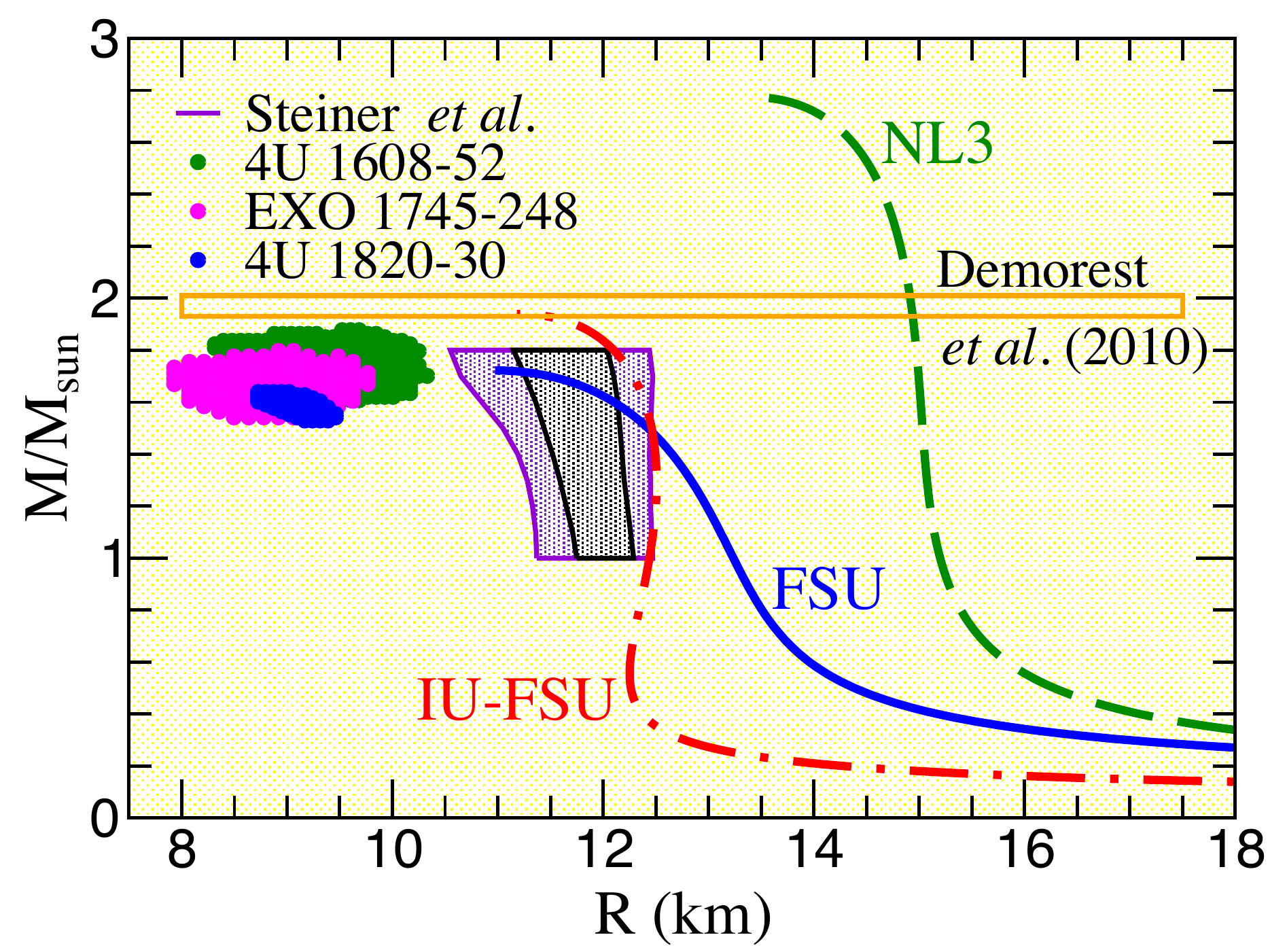}
 \vspace{-0.2cm}
 \caption{(Color online) {\sl Mass-vs-Radius} relation predicted by 
  the three relativistic mean-field models\,\cite{Fattoyev:2010mx}. The
  observational data that suggest very small stellar radii represent 
  1$\sigma$ confidence contours for the three neutron stars reported 
   in Ref.\,\cite{Ozel:2010fw}. The two shaded areas that suggest larger radii
  are 1$\sigma$ and 2$\sigma$ contours extracted from the analysis of 
  Ref.\,\cite{Steiner:2010fz}. The seminal measurement of the heaviest
  neutron star by Demorest et al., is also indicated in the 
  figure\,\cite{Demorest:2010bx}.}
 \label{Fig2}
\end{center}
\end{figure}

Neutron stars satisfy the Tolman-Oppenheimer-Volkoff (TOV) equations, which 
are the extension of Newton's laws to the domain of general relativity. The TOV 
equations may be expressed as a coupled set of first-order differential equations 
of the following form:
 \begin{eqnarray}
   && \frac{dP}{dr}=-G\,\frac{{\cal E}(r)M(r)}{r^{2}}
         \left[1+\frac{P(r)}{{\cal E}(r)}\right]
         \left[1+\frac{4\pi r^{3}P(r)}{M(r)}\right]
         \left[1-\frac{2GM(r)}{r}\right]^{-1} \;,
         \label{TOVa}\\
   && \frac{dM}{dr}=4\pi r^{2}{\cal E}(r)\;,
         \label{TOVb}
 \label{TOV}
\end{eqnarray}
where $G$ is Newton's gravitational constant and $P(r)$, ${\cal E}(r)$, and $M(r)$ 
represent the pressure, energy density, and enclosed-mass profiles of the star, 
respectively. Note that the  three terms enclosed in square brackets in Eq.~(\ref{TOVa}) 
are of general-relativistic origin. Notably, the only input that neutron stars are sensitive 
to is the equation of state (EOS), namely, $P\!=\!P({\cal E})$. In Fig.\,\ref{Fig2} we display 
{\sl mass-{\sl vs}-radius} relations as predicted by three relativistic mean-field 
models\,\cite{Fattoyev:2010mx}. Surprisingly, all three 
models---NL3\,\cite{Lalazissis:1996rd, Lalazissis:1999},
FSU\,\cite{Todd-Rutel:2005fa}, and IU-FSU\,\cite{Fattoyev:2010mx},
are able to accurately reproduce a variety of ground-state observables throughout 
the nuclear chart. Yet, the predictions displayed in Fig.\,\ref{Fig2} are significantly 
different. A central goal of this contribution is to identify critical laboratory observables 
that may be used to constrain the structure, dynamics, and composition of neutron stars.

The starting point for the calculation of both nuclear and neutron-star structure is 
a relativistic energy density functional characterized by the interacting Lagrangian 
density of Ref.~\cite{Mueller:1996pm} supplemented by an isoscalar-isovector term
first introduced in Ref.~\cite{Horowitz:2000xj}. That is,
\begin{eqnarray}
{\mathscr L}_{\rm int} &=&
\bar\psi \left[g_{\rm s}\phi   \!-\! 
         \left(g_{\rm v}V_\mu  \!+\!
    \frac{g_{\rho}}{2}{\mbox{\boldmath $\tau$}}\cdot{\bf b}_{\mu} 
                               \!+\!    
    \frac{e}{2}(1\!+\!\tau_{3})A_{\mu}\right)\gamma^{\mu}
         \right]\psi \nonumber \\
                   &-& 
    \frac{\kappa}{3!} (g_{\rm s}\phi)^3 \!-\!
    \frac{\lambda}{4!}(g_{\rm s}\phi)^4 \!+\!
    \frac{\zeta}{4!}   g_{\rm v}^4(V_{\mu}V^\mu)^2 +
   \Lambda_{\rm v}\Big(g_{\rho}^{2}\,{\bf b}_{\mu}\cdot{\bf b}^{\mu}\Big)
                           \Big(g_{\rm v}^{2}V_{\nu}V^{\nu}\Big)\;.
 \label{LDensity}
\end{eqnarray}
The Lagrangian density includes an isodoublet nucleon field ($\psi$)
interacting via the exchange of two isoscalar mesons, a scalar
($\phi$) and a vector ($V^{\mu}$), one isovector meson ($b^{\mu}$),
and the photon ($A^{\mu}$)~\cite{Serot:1984ey,Serot:1997xg}. In
addition to meson-nucleon interactions, the Lagrangian density is
supplemented by four nonlinear meson interactions with coupling
constants ($\kappa$, $\lambda$, $\zeta$, and $\Lambda_{\rm v}$)  
that are included primarily to soften the equation of state of both
symmetric nuclear matter and pure neutron matter. For a detailed
discussion on the impact of these terms on various quantities of
theoretical, experimental, and observational interest see
Ref.\,\cite{Piekarewicz:2007dx}.

\begin{figure}[h]
\begin{center}
 \includegraphics[width=8cm,height=6cm]{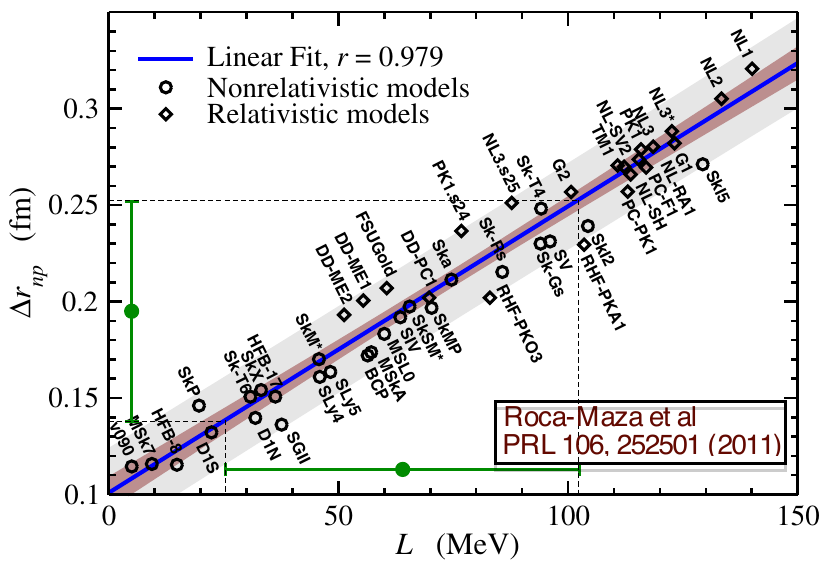}
  \hspace{0.1cm}
 \includegraphics[width=6cm,height=6cm]{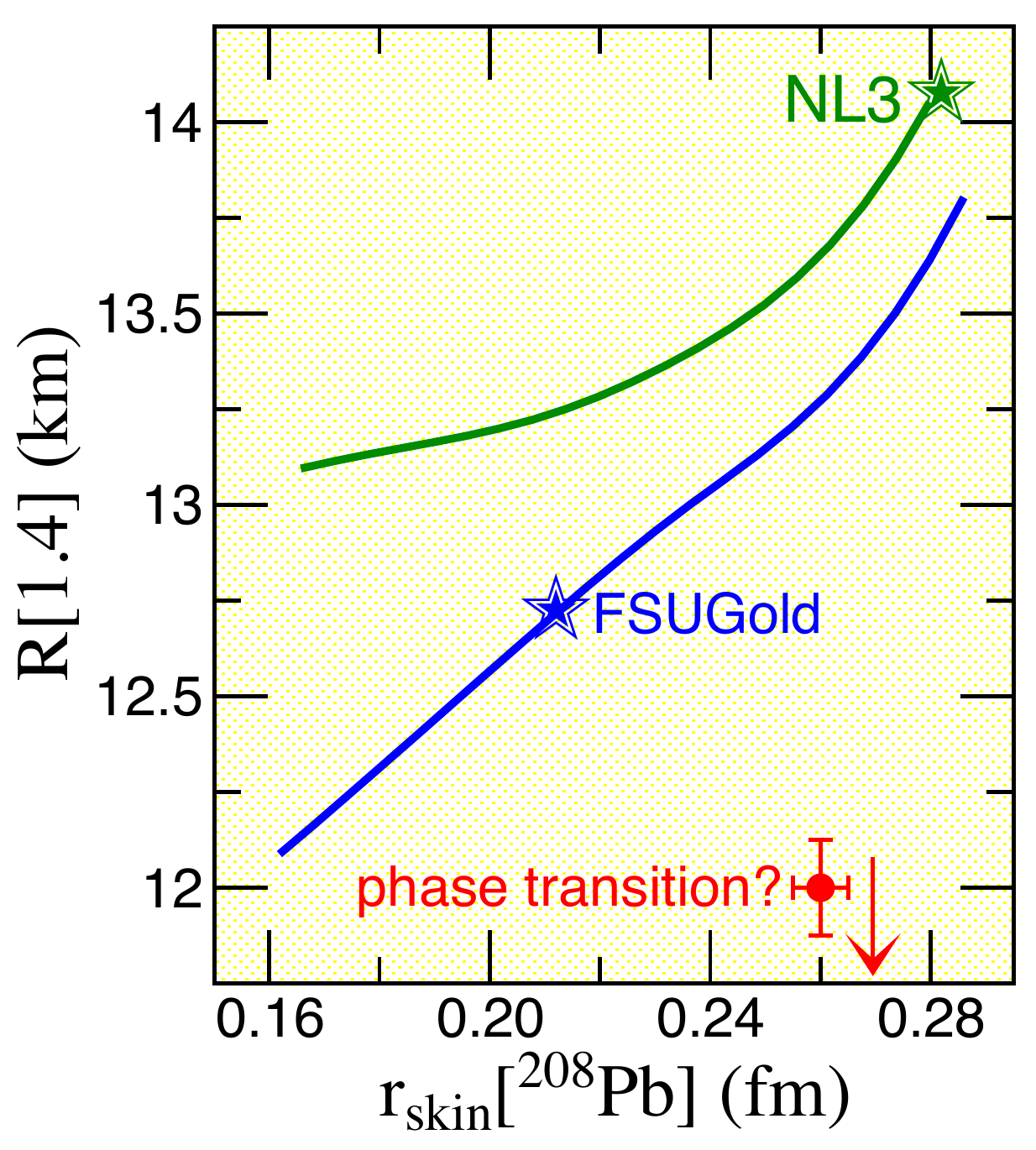}
 \vspace{-0.2cm}
 \caption{(Color online) The left-hand panel displays the correlation 
   between the neutron-skin of ${}^{208}$Pb and the slope of the 
   symmetry energy for a variety of nonrelativistic and relativistic 
   models\,\cite{RocaMaza:2011pm}. The right-hand panel shows the 
   correlation between the neutron-skin of ${}^{208}$Pb and the radius 
   of a $1.4 \,{\rm M}_{\odot}$ neutron star for two relativistic
   mean-field models.}      
 \label{Fig3}
\end{center}
\end{figure}

Laboratory experiments may play a critical role in constraining the 
size of neutron stars because stellar radii are controlled by the 
density dependence of the symmetry energy in the immediate 
vicinity of nuclear-matter saturation density\,\cite{Lattimer:2006xb}.  
Recall that the symmetry energy may be viewed as the difference 
in the energy between pure neutron matter and symmetric nuclear 
matter.  A particularly critical property of the symmetry energy that
is poorly constrained is its slope at saturation density---a quantity 
customarily denoted by $L$ and closely related to the pressure of 
pure neutron matter\,\cite{Piekarewicz:2008nh}. Although $L$ is 
not directly observable, it is strongly correlated to the thickness of 
the neutron skin of heavy nuclei\,\cite{Brown:2000,Furnstahl:2001un}.  
As indicated in the left-hand panel of Fig.\,\ref{Fig3}, the thickness 
of the neutron skin depends sensitively on the pressure of neutron-rich 
matter: {\sl the greater the pressure the thicker the neutron skin}. And 
it is exactly this same pressure that supports neutron stars against 
gravitational collapse. Thus, as indicated in the right-hand panel of 
Fig.\,\ref{Fig3}, models with thicker neutron skins often produce neutron 
stars with larger radii\,\cite{Horowitz:2000xj,Horowitz:2001ya}. Thus, it is 
possible to study ``data-to-data'' relations between the neutron-rich skin of 
a heavy nucleus and the radius of a neutron star.

\section{Neutron Skins: The Lead Radius Experiment (PREX)}
\label{PREX} 

The successfully commissioned Lead Radius Experiment has provided 
the first model-independent evidence of the existence of a significant 
neutron skin in ${}^{208}$Pb\,\cite{Abrahamyan:2012gp,Horowitz:2012tj}.
Parity violation at low momentum transfers is particularly sensitive to the 
neutron distribution because the neutral weak-vector boson ($Z^0$) couples
preferentially to the neutrons in the target~\cite{Donnelly:1989qs}.
Although PREX achieved the systematic control required to perform this
challenging experiment, unforeseen technical problems resulted in time
losses that significantly compromised the statistical accuracy of the
measurement. This resulted in the following value for the neutron-skin
thickness of ${}^{208}$Pb\,\cite{Abrahamyan:2012gp,Horowitz:2012tj}:
\begin{equation}
r_{\rm skin}\!=\!R_{n}\!-\!R_{p}\!=\!{0.33}^{+0.16}_{-0.18}~{\rm fm}.
\end{equation}
Given that the determination of the neutron radius of a heavy nucleus
is a problem of fundamental importance with far reaching implications
in areas as diverse as nuclear
structure~\cite{Brown:2000,Furnstahl:2001un,Danielewicz:2003dd,
Centelles:2008vu,Centelles:2010qh}, atomic parity
violation~\cite{Pollock:1992mv,Sil:2005tg}, heavy-ion
collisions~\cite{Tsang:2004zz,Chen:2004si,Steiner:2005rd,
Shetty:2007zg,Tsang:2008fd}, and neutron-star
structure~\cite{Horowitz:2000xj,Horowitz:2001ya,Horowitz:2002mb,
Carriere:2002bx,Steiner:2004fi,Li:2005sr,Fattoyev:2010tb}, the PREX
collaboration has made a successful proposal for additional beam time
so that the original 1\% goal (or $\!\pm0.05$\,fm) may be 
attained\,\cite{PREXII:2012}. While the scientific case for such a 
critical experiment remains strong, the search for additional physical
observables that may be both readily accessible and strongly
correlated to the neutron skin (and thus also to $L$) is a worthwhile
enterprise. It is precisely the exploration of such a correlation
between the {\sl electric dipole polarizability} and the neutron-skin
thickness of ${}^{208}$Pb that is at the center of the next section.

\section{Pygmies and Giant Resonances}
\label{PygmiesGiants} 

A promising complementary approach to the parity-violating program
relies on the electromagnetic excitation of the electric dipole
mode~\cite{Harakeh:2001}.  For this mode of excitation---perceived 
as a collective oscillation of neutrons against protons---the symmetry 
energy acts as the restoring force. In particular, models with a soft 
symmetry energy predict large values for the symmetry energy at the 
densities of relevance to the excitation of this mode. As a consequence, 
the stronger restoring force of the softer models generates a dipole 
response that is both hardened and quenched relative to its stiffer 
counterparts.  In particular, the {\sl inverse} energy-weighted sum, 
which is directly proportional to the electric dipole polarizability 
$\alpha_{\raisebox{-1pt}{\tiny D}}$, is highly sensitive to the density 
dependence of the symmetry energy\,\cite{Piekarewicz:2010fa}. 
This sensitivity suggests the existence of the following interesting 
correlation in heavy nuclei: {\sl the larger $r_{\rm skin}$
the larger~$\alpha_{\raisebox{-1pt}{\tiny D}}$}.

\begin{figure}[h]
\begin{center}
 \includegraphics[height=6cm]{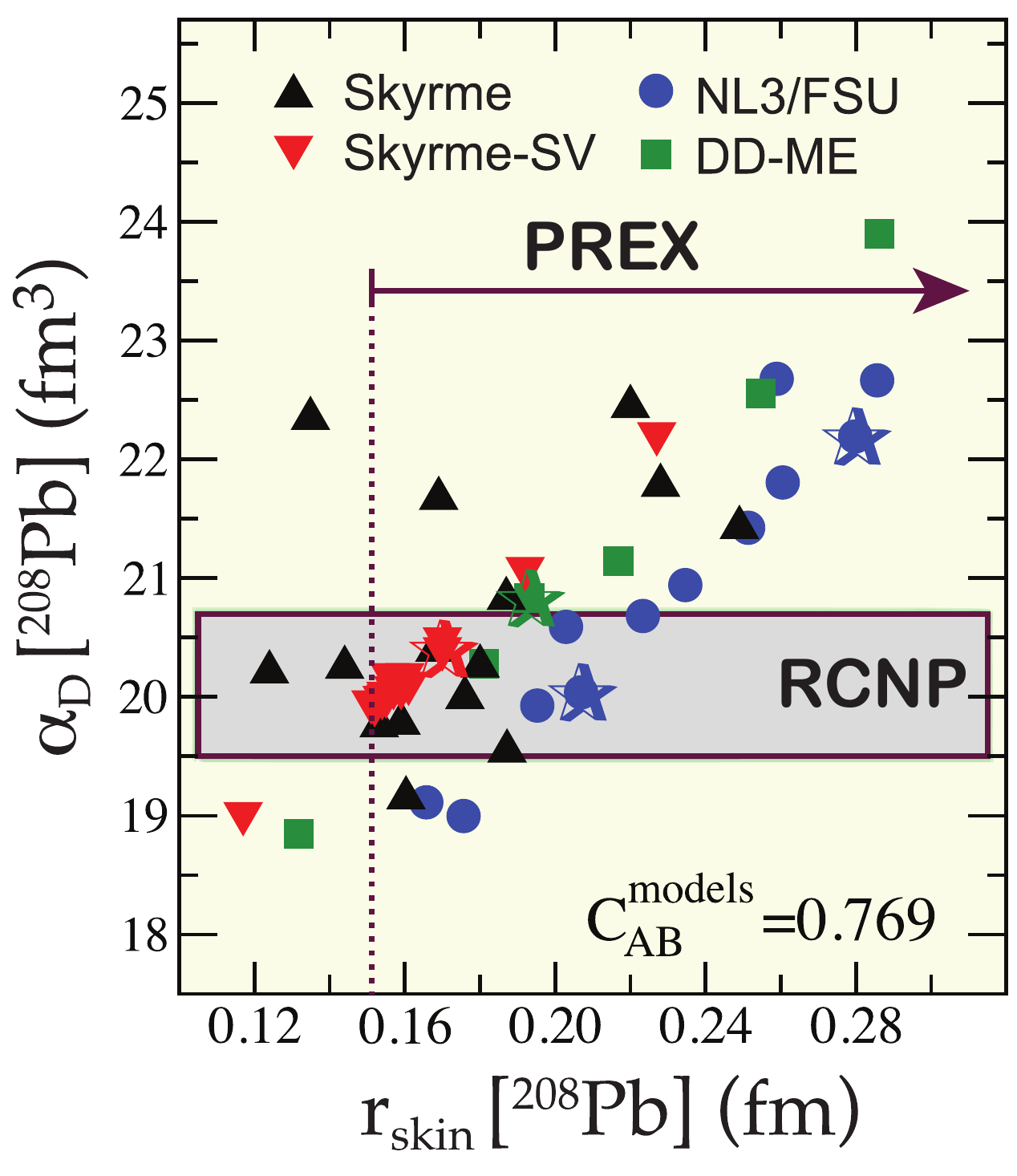}
  \hspace{0.2cm}
 \includegraphics[height=6cm]{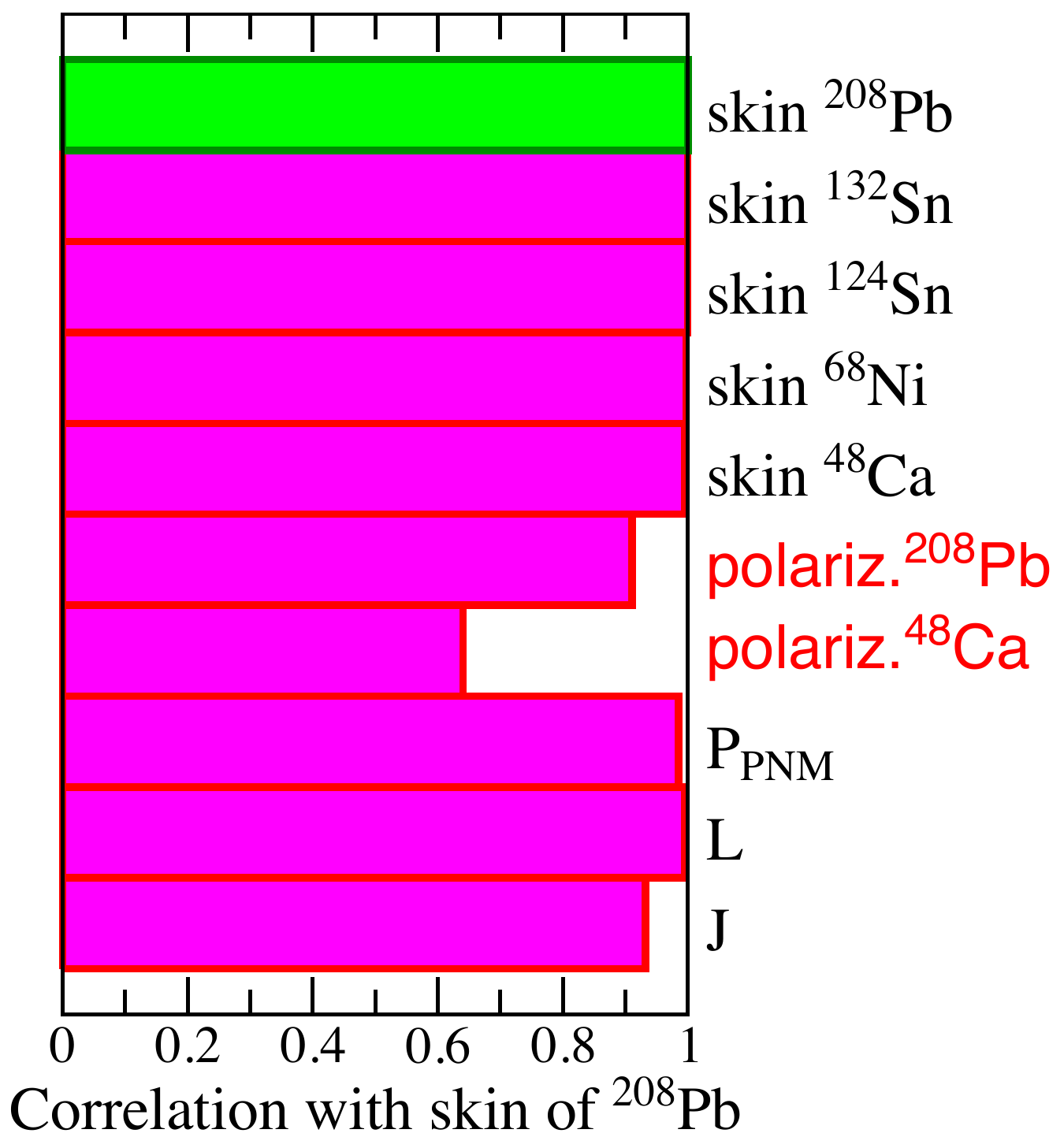}
 \vspace{-0.2cm}
 \caption{(Color online) Predictions from a variety of nuclear models
  for the electric dipole polarizability and neutron-skin thickness 
  of ${}^{208}$Pb are shown on the left-hand side of the figure. Also
  shown are constrains on the neutron-skin thickness from 
  PREX~\cite{Abrahamyan:2012gp,Horowitz:2012tj} and on the 
  dipole polarizability from RCNP~\cite{Tamii:2011pv,
  Poltoratska:2012nf}. On the right-hand side of the figure we
  show  correlation coefficients between the neutron-skin thickness 
  of ${}^{208}$Pb and several observables as obtained from a
  covariance analysis based on the FSU 
  interaction~\cite{Fattoyev:2012rm}.}
\label{Fig4}
\end{center}
\end{figure}

To test the validity of this correlation we display on the left-hand panel 
of Fig.\,\ref{Fig4} $\alpha_{\raisebox{-1pt}{\tiny D}}$ in ${}^{208}{\rm Pb}$ 
as a function of its corresponding neutron-skin as predicted by a large 
number (48) of nuclear-structure models\,\cite{Piekarewicz:2012pp}.   
From the distribution of electric dipole strength 
($R_{\raisebox{-1pt}{\tiny E1}}$) the dipole polarizability is readily 
extracted from the inverse energy-weighted sum. That is,
\begin{equation}
 \alpha_{\raisebox{-1pt}{\tiny D}} = \frac{8\pi}{9}e^{2}
  \int_{0}^{\infty}\!\omega^{-1} 
   R_{\raisebox{-1pt}{\tiny E1}} (\omega)\,d\omega \;.
\label{AlphaD}
\end{equation}
At first glance a clear (positive) correlation between the dipole
polarizability and the neutron skin is discerned. However, on 
closer examination one observes a significant scatter in the
results---especially in the case of the standard Skyrme forces
(denoted by the black triangles).  In particular, by including the
predictions from all the 48 models under consideration, a correlation
coefficient of 0.77 was obtained.  Also shown in the figure are
experimental constraints imposed from PREX and the recent
high-resolution measurement of $\alpha_{\raisebox{-1pt}{\tiny D}}$
in $^{208}$Pb\,\cite{Tamii:2011pv,Poltoratska:2012nf}.  By imposing
these recent experimental constraints, several of the
models---especially those with either a very soft or very stiff
symmetry energy---may already be ruled out.  

However, to establish how the dipole polarizability may provide a
unique constraint on the neutron-skin thickness of neutron-rich nuclei
and other isovector observables we display on the right-hand panel of
Fig.\,\ref{Fig4} correlation coefficients computed using a single
underlying model, namely, FSU\,\cite{Todd-Rutel:2005fa}. For details
on the implementation of the required covariance analysis we refer the
reader to
Refs.\,\cite{Reinhard:2010wz,Fattoyev:2011ns,Fattoyev:2012rm}. According
to the model, an accurate measurement of the neutron skin-thickness in
${}^{208}{\rm Pb}$ significantly constrains the neutron skin on a
variety of other neutron-rich nuclei. Moreover, the correlation
coefficient between the neutron skin and
$\alpha_{\raisebox{-1pt}{\tiny D}}$ in ${}^{208}{\rm Pb}$ is very
large (of about 0.9). This suggests that a multi-prong approach
consisting of combined measurements of both neutron skins and
$\alpha_{\raisebox{-1pt}{\tiny D}}$---ideally on a variety of
nuclei---should significantly constrain the isovector sector of the
nuclear energy density functional as well as the EOS of neutron-rich
matter.

Naturally, a more stringent constrain on the isovector sector of the
nuclear density functional is expected to emerge along an isotopic
chain as the nucleus develops a neutron-rich skin. Concomitant with 
the development of a neutron skin one expects the emergence of low 
energy dipole strength---the so-called {\sl pygmy dipole
resonance}\,\cite{Suzuki:1990,VanIsacker:1992,Hamamoto:1996,
Hamamoto:1998,Vretenar:2000yy,Vretenar:2001hs,Paar:2004gr}.  
Thus, it has been suggested that the pygmy dipole resonance
(PDR)---speculated to be an excitation of the neutron-rich skin
against the isospin symmetric core---may be used as a constraint 
on the neutron skin\,\cite{Piekarewicz:2006ip}. 

\begin{figure}[h]
\begin{center}
 \includegraphics[height=6cm]{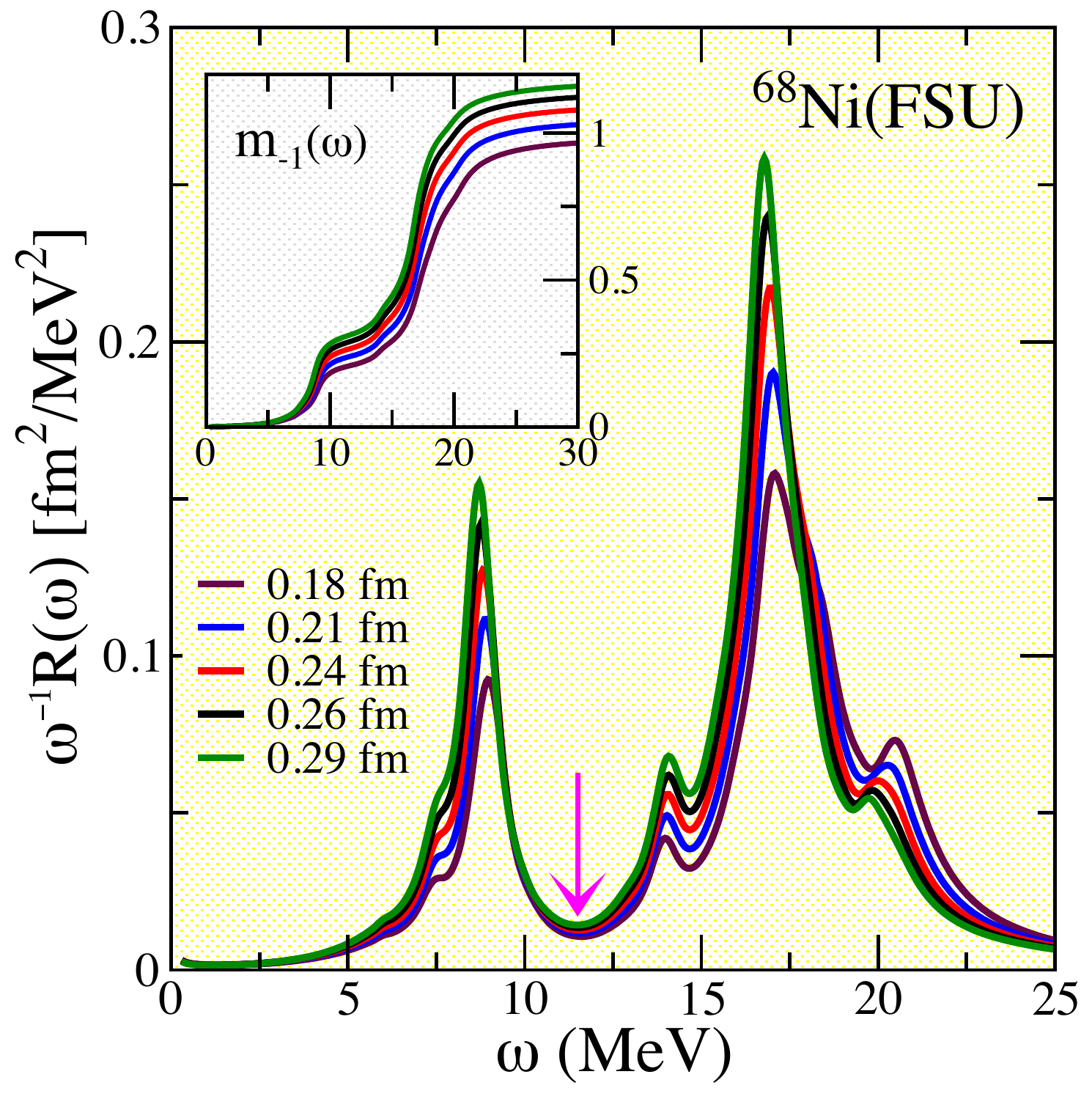}
  \hspace{0.1cm}
 \includegraphics[height=6cm]{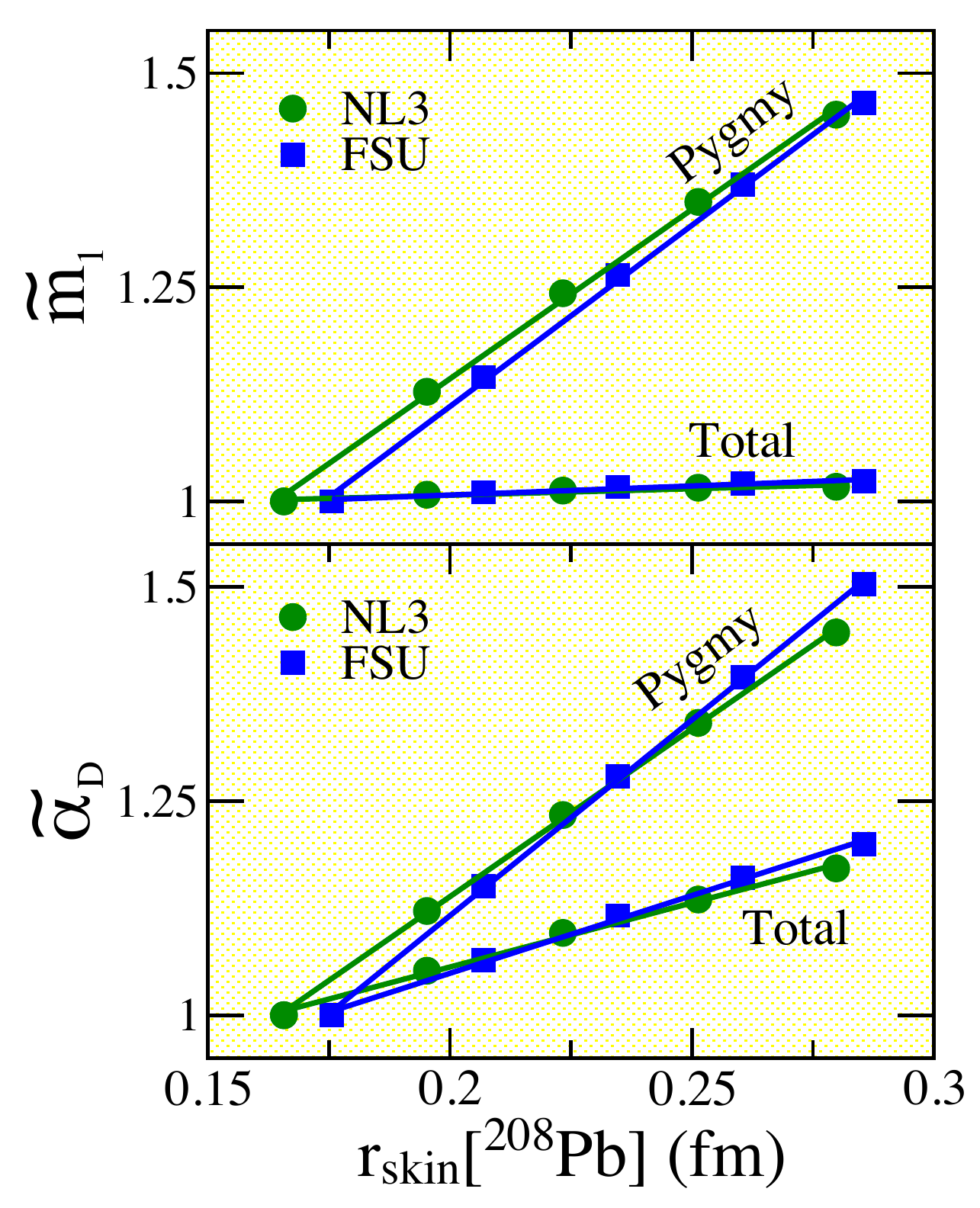}
 \vspace{-0.2cm}
 \caption{(Color online) The inverse energy weighted dipole response
   in ${}^{68}$Ni computed with the FSU family of effective interactions
   is shown on the left-hand side of the figure. The inset displays the 
   running sum. The arrow
   indicates the (ad-hoc) energy at which the low-energy (pygmy)
   response is separated from the high-energy (giant) response. 
   On the right-hand side the fractional change in the energy weighted 
   sum and dipole polarizability for ${}^{68}$Ni are displayed as a
   function of the neutron-skin thickness of ${}^{208}$Pb. See 
   Ref.\,\cite{Piekarewicz:2010fa} for more details.}
\label{Fig5}
\end{center}
\end{figure}
In particular, the fraction of the {\sl inverse energy weighted sum
rule} concentrated on the low-energy region appears to be sensitive
to the neutron skin of neutron-rich nuclei. The inverse energy weighted 
response $\omega^{-1} R(\omega)$ is displayed on the left-hand panel 
of Fig.\,\ref{Fig5} for the neutron-rich nucleus ${}^{68}$Ni.
Given that the $\omega^{-1}$ factor enhances the 
low-energy part of response, the Pygmy resonance accounts for a 
significant fraction (of about 20-25\%) of $m_{-1}$ (which is proportional 
to $\alpha_{\raisebox{-1pt}{\tiny D}}$). Pictorially, this behavior is best 
illustrated in the inset of Fig.\,\ref{Fig5} which displays the {\sl ``running''}
$m_{-1}(\omega)$ sum. The inset provides a clear indication that both 
the total $m_{-1}$ moment as well as the fraction contained in the Pygmy 
resonance are highly sensitive to the neutron-skin thickness of ${}^{208}$Pb.  
To heighten this sensitivity we display on the right-hand panel of
Fig.\,\ref{Fig5} the {\sl fractional change} in both the total and
Pygmy contributions to the $m_{1}$ moment ({\sl i.e.,} the energy weighted 
sum rule) and to the dipole
polarizability $\alpha_{D}$ as a function of the neutron skin of
${}^{208}$Pb (we denote these fractional changes with a {\sl
``tilde''} in the figure). These results illustrate the strong
correlation between the neutron skin and $\alpha_{D}$ and establish
how a combined measurement of these laboratory observables will be of
vital importance in constraining the isovector sector of the nuclear
density functional.

\section{Conclusions}
\label{conclusions} 
 
Measurements of neutron radii provide important constraints on the
isovector sector of nuclear density functionals and offer vital
guidance in areas as diverse as atomic parity violation, heavy-ion
collisions, and neutron-star structure. In this contribution we
examined the possibility of using the quintessential nuclear
mode---the isovector dipole resonance---as a promising 
complementary observable.
For this mode of excitation in which protons oscillate coherently
against neutrons, the symmetry energy acts as its restoring force.
Thus, models with a soft symmetry energy predict large values for the
symmetry energy at the densities of relevance to the excitation of
this mode. As a consequence, softer models generates a dipole response
that is both hardened and quenched relative to the stiffer
models. However, being protected by the Thomas-Reiche-Kuhn sum rule,
the energy weighted sum rule is largely insensitive to this behavior. In
contrast, for the inverse energy-weighted sum---which is
directly proportional to the electric dipole polarizability
$\alpha_{\raisebox{-1pt}{\tiny D}}$---the quenching and hardening act
in tandem.  Thus, models with a soft symmetry energy predict smaller
values of $\alpha_{\raisebox{-1pt}{\tiny D}}$ than their stiffer
counterparts. This results in a powerful ``data-to-data'' relation:
{\sl the smaller $\alpha_{\raisebox{-1pt}{\tiny D}}$ the thinner the
neutron skin}. Moreover, we saw that a significant amount of 
electric dipole strength is concentrated in the low-energy fragment;
the so-called Pygmy resonance.

In summary, motivated by two seminal
experiments\,\cite{Abrahamyan:2012gp,Tamii:2011pv}, we examined
possible correlations between the electric dipole polarizability and
the neutron skin of neutron-rich nuclei. The neutron-skin thickness of
a heavy nucleus is a quantity of critical importance for our
understanding of a variety of nuclear and astrophysical phenomena.  In
particular, the neutron-skin thickness of $^{208}$Pb can provide
stringent constrains on the density dependence of the symmetry energy
which, in turn, has a strong impact on the structure, dynamics, and
composition of neutron stars.  We conclude that precise measurements
of neutron skins and $\alpha_{\raisebox{-1pt}{\tiny D}}$---ideally on
a variety of nuclei--- should significantly constrain the isovector
sector of the nuclear energy density functional and will provide
critical insights into the nature of neutron-rich matter.

\section*{Acknowledgments}
  This work was supported in part by grant DE-FD05-92ER40750
  from the Department of Energy.

\end{document}